\documentclass{elsarticle}
\usepackage{amsmath}
\usepackage{mathrsfs}
\usepackage{bbm}
\usepackage{graphicx}
\usepackage{calligra}
\usepackage{quotmark}
\usepackage{longtable}
\usepackage{cases} 
\makeatletter
\def\ps@pprintTitle{%
 \let\@oddhead\@empty
 \let\@evenhead\@empty
 \def\@oddfoot{\centerline{\thepage}}%
 \let\@evenfoot\@oddfoot}
\makeatother
\begin{document}
\begin{frontmatter}
\begin{center}
{\Large{\bf Reproducing Kernel Functions: A general framework for Discrete Variable Representation}}
\end{center}
\begin{center}
{Hamse Y. Mussa}
\end{center}
\begin{center}
{\it Unilever Centre for Molecular Sciences Informatics, Department of Chemistry, University of Cambridge, Lensfield Road, Cambridge CB2 1EW, UK}
\end{center}
\begin{center}
\end{center}
\begin{abstract}
 Since its introduction, the Discrete Variable Representation (DVR) basis set has become an 
 invaluable representation of state vectors and Hermitian operators in non-relativistic quantum 
 dynamics and spectroscopy calculations. On the other hand reproducing kernel (positive definite) functions 
 have been widely employed for a long time to a wide variety of disciplines: detection and estimation problems in signal processing;  data analysis in statistics; generating observational models in machine learning; solving inverse problems 
 in geophysics and tomography in general; and in quantum mechanics. 
 
 In this article it was demonstrated that, starting with the axiomatic definition of DVR provided 
 by Littlejohn {\it et al} \cite{Littlejohn}, it is possible to show that the space upon which  
 the projection operator defined in ref \cite{Littlejohn} projects is a Reproducing Kernel 
 Hilbert Space (RKHS) whose associated reproducing kernel function can be used to generate DVR points 
 and their corresponding DVR functions on any domain manifold (curved or not). It is illustrated how, with this idea, 
 one may be able to `neatly' address the long-standing challenge of building multidimensional DVR basis 
 functions defined on curved manifolds.
\end{abstract}

\end{frontmatter}
\section{Introduction}

 Studies in basic and ordinary (non-relativistic) quantum dynamics and spectroscopy have made great 
 progress in the past few decades. Arguably, expressing the relevant higher dimensional dynamic operators 
 and state vectors in the so-called direct product DVR basis sets \cite{Light,Light-carr} has been a major 
 contributor to this progress. However, the configuration manifolds on which those states and Hermitian 
 operators are described are in general not Euclidean \cite{Sutcliffe,Seran,Xaus}. In those scenarios, it was 
 pointed out that proper multidimensional DVR basis set functions defined over curved manifolds cannot, 
 in general, be written in a direct product form \cite{Light-carr}(and references therein). Constructing 
 the correct DVR basis set on curved configuration spaces has been a challenging problem since the mid 1980's,
 when Light and co-workers demonstrated the enormous potential in using DVR techniques for solving the appropriate 
 nuclear motion Schrodinger equations for polyatomic molecules \cite{Light}. 
 
 To our knowledge, there has not been much progress on the issue until recently when a few promising 
 papers appeared \cite{Littlejohn,Dawes,Yu,Tennor}. The first of those was presented by Littlejohn 
 {\it et al}. Here in this article one of the main objectives is to revisit the excellent work presented in 
 that paper. We `restate' the core idea in ref \cite{Littlejohn} as a definition and then demonstrate that 
 the DVR concept can be connected with the theory of reproducing kernel functions. It is then illustrated how 
 this observation may lead to a `neat' solution to the aforementioned outstanding problem of building 
 Multidimensional Non-direct Product (MNPDVR) basis sets on metric and differentiable manifolds. In the 
 following section, we give a brief overview of the DVR concept. In Section 3 the central idea of the paper 
 is presented. The final section gives some concluding remarks. Note that all spaces, variables, matrices 
 and manifolds considered here are real unless stated otherwise. Also note that the two statements `flat 
 manifold' and `Euclidean manifold' are loosely used interchangeably.

\section{Brief overview of Direct Product DVR} 

 Although in this paper a familiarity with the DVR scheme \cite{Light, Light-carr,Harris, Dickenson, szlay,
 Manolopoulos,Muckerman,Turker,Miller,Wu,Fulton} and its employment in practice is assumed, a brief outline 
 of the basics of the DVR concept is provided --- albeit, an introduction tailored mainly to set the scene 
 for us to introduce the reproducing kernel theory into the DVR techniques. 
 
 In nuclear, atomic (and, of course, molecular) levels it is acceptable to ignore the effects of 
 gravity (and quantum gravity for that matter). In other words, the physical space is considered 
 a homogeneous, isotropic, and continuum entity that is subject to the laws of Euclidean geometry. 
 One important implication of these assumptions is that the spectrum of a position operator defined 
 on the Euclidean `manifold' is continuous. It also means if the molecular system consists of $\mathit{L}$ 
 atoms, it is unavoidable to consider the configuration space of the system as a Euclidean {\it 3}$\mathit{L}$-dimensional 
 differentiable metric manifold, $\mathbbm{M}$. In this setting, in global coordinate system descriptions, 
 each point ${\alpha}$ in $\mathbbm{M}$ is represented by a set of $\mathit{3L(=d)}$ coordinates, which are 
 denoted here by $x^{(i=1,....d)}_{\alpha}$. A consequence of the physical system's `arena' being a flat
 $\mathit{d}$-dimensional manifold is that all the $\mathit{d}$ components of the position operator 
 $\hat{x}^{(i=1,....d)}$ are mutually commutative, {\it i.e.}, they are simultaneously diagonalisable 
 on $\mathbbm{M}$.
 
 Thus, ideally, the so-called `generalised eigenfunction' of (say) ${\hat{x}}^{(1)}$ at $x^{(1)}_{\alpha}$ 
 can be seen as a Dirac's tempered distribution, $\delta(x^{(1)}-x^{(1)}_{\alpha})$, and the eigenfunction of $\hat{x}^{(1:d)}$ 
 can then be expressed as $\delta(x^{1:d}-x^{1:d}_{\alpha}) = \Pi_{i=1}^{d}\delta(x^{(i)}- x^{(i)}_{\alpha})$ in Cartesian coordinates.
 Obviously, these `functions' do not belong to $\mathscr{L}_{(\mathbbm{M})}^{2}$ --- an infinite-dimensional 
 space whose elements are square-integrable functions on $\mathbbm{M}$, which in the rest of the paper we 
 denote $\mathscr{H}$. However, given a set of $N$ (finite number) orthonormal functions $\{\psi_{j}(x^{i})\}_{j=1}^{N}\in\mathscr{H}$, 
 one can (in principle) for $\hat{x}^{(1:d)}$ yield $N$ eigenvalues $\{x^{(1:d)}_{\alpha}\}^{N}_{\alpha=1}$ 
 in $\mathbbm{M}$ and their corresponding eigenfunctions $\{u_{\alpha}(x^{1:d})\}^{N}_{\alpha=1}$ in $\mathscr{H}$, 
 such that $u_{\alpha}(x^{1:d})=\Pi_{i=1}^{d}\rho^{i}_{\alpha}(x^{i})$; with $\rho^{i}_{\alpha}(x^{i})$ being 
 the eigenfunction associated with $x^{i}_{\alpha}$ and an element of $\mathscr{H}$ \cite{Harris}, that in the 
 $\displaystyle\lim_{N\to\infty}$ becomes a Dirac delta function which, as pointed out, earlier does 
 not belong to $\mathscr{H}$. 
 
 $\{x^{(1:d)}_{\alpha}\}^{N}_{\alpha=1}$ and $\{u_{\alpha}(x^{i:d})\}^{N}_{\alpha=1}$ 
 are the so-called DVR points and basis set, respectively \cite{Harris,Golub}. The DVR basis functions 
 expressed in the above composite form is an example of what is commonly termed the direct product DVR basis 
 set \cite{Light,Light-carr}. Computing $\{x^{(1:d)}_{\alpha},u_{\alpha}(x^{1:d})\}$ on $\mathbbm{M}$ is a 
 straightforward matter. One simple way of achieving this is to express (say) $\hat{x}^{(i)}$ in $\{\psi_{j}(x^{i})\}_{j=1}^{N}$ which 
 yields a $N\times N$ matrix for this particular position operator component. The eigen-pairs of the matrix are the 
 DVR points and DVR eigenvectors of $\hat{x}^{(i)}$, respectively \cite{Harris,Golub}. 
 For later reference, we provide below a formal form of the eigenfunction for (as an example) $\hat{x}^{(i)}$ represented in $\{\psi_{j}(x^{i})\}_{j=1}^{N}$:
 \begin{equation}
  \rho^{i}_{\alpha}(x^{i})=\sum_{j}^{N}\mathit{R}_{\alpha j}\psi_{j}(x^{(i)})
 \end{equation}
 where 
 \begin{equation}
 \mathit{R}_{\alpha j}=\int^{\infty}_{-\infty}dx^{(i)}\psi_{j}(x^{(i)})\rho^{i}_{\alpha}(x^{(i)})
 \end{equation}
 denotes the matrix elements of the unitary transformation matrix \cite{Dickenson,Golub}. If the quadrature
 $\int^{\infty}_{-\infty}dx^{(i)}\psi_{j}(x^{(i)})\rho^{i}_{\alpha}(x^{(i)})=\sum_{\beta}^{N}\psi_{j}(x^{(i)}_{\beta})\rho^{i}_{\alpha}(x^{(i)}_{\beta})$ exists, then 
 \begin{equation} 
 \mathit{R}_{\alpha j}=\int^{\infty}_{-\infty}dx^{(i)}\psi_{j}(x^{(i)})\rho^{i}_{\alpha}(x^{(i)})=c_{\alpha}\psi_{j}(x^{(i)}_{\alpha}) 
 \end{equation}
 because by definition $\rho^{i}_{\alpha}(x^{(i)}_{\beta})\propto\delta_{\alpha\beta}$ \cite{Dickenson}; 
 and $c_{\alpha}\in\mathscr{R}$. (Note that the superscripts in $\rho^{i}$, $x^{(1:d)}$, {\it etc}., are
 merely for labelling.)
 
 In summary, superficially, one may view DVR basis functions as the representation of physically 
 realizable position states. In that case, finding a DVR basis function is mathematically equivalent 
 to obtaining bounded functional of $\mathscr{D}\subseteq\mathscr{L}_{(\mathbbm{M})}^{2}$ 
 that approximates (in the $\displaystyle\lim_{N\to\infty}$) a Dirac delta functional 
 of $\mathscr{L}_{(\mathbbm{M})}^{2}$. A potent tool for this is the rich theory of operators and 
 functionals on Hilbert spaces \cite{Glazman,Natingale,Yoshika}. However, this approach would be 
 too abstract for the purpose of this paper. 

 At any rate, for both physical insight and calculation expedience, in molecular physics 
 the internal motions ({\it i.e.}, shape defining motions) of polyatomic molecules is described in 
 a $\bar{{\it d}}$-dimensional surface embedded in $\mathbbm{M}$, where $\bar{\mathit{d}}<\mathit{ d}$ 
 \cite{Sutcliffe,Seran,Xaus,Bramley}. This is achieved by introducing some appropriate constraints. 
 The imposition of these constraints turns the physical system's configuration space from Euclidean 
 to Riemannian \cite{Seran, Xaus}, denoted here by $\mathbbm{M}_{c}$. In simple terms, the $\bar{{\it d}}$-dimensional 
 subspace embedded in the original and larger $\mathit{d}$-dimensional Euclidean configuration space does not 
 have any internal/inherent Euclidean metric (hence it is curved), but this subspace looks like a Euclidean 
 space locally. Unfortunately, on the Riemannian manifold the molecular system can only intrinsically be 
 described in curvilinear co-ordinate systems, and besides there is no single internal curvilinear co-ordinate
 system which can cover $\mathbbm{M}_{c}$ globally. This means there are some of the embedded molecular configurations,
 such that their curvilinear coordinates are not defined. In other words, there is no co-ordinate system on this 
 curved manifold that can provide a unique label for each constrained molecular configuration. In the following 
 discussion, in lieu of explicitly reciting each time the full rigmarole about the mapping relationship between 
 co-ordinate systems and manifolds, the statement `a function is defined over $\mathbbm{M}_{c}$' means that the 
 function is defined over a patch in the manifold where the co-ordinates are valid, see Refs.\cite{Schuts,Tod,Tensor}. 
 This is also discussed in the chemical physics literature, see Refs.\cite{Sutcliffe,Seran}for more accessible details.

 Owing to the aforementioned constraints, it can no longer guaranteed that all $\bar{\mathit{d}}$ 
 components of the co-ordinate operators defined in these curvilinear co-ordinates are mutually 
 commutative. In other words, the components may not be diagonalisable simultaneously on $\mathbbm{M}_{c}$. 
 Thus, employing a direct product DVR approach to represent higher dimensional Hermitian operators defined 
 in the curvilinear co-ordinates in $\mathbbm{M}_{c}$ might lead to erroneous estimates of the observables. 
 
 As briefly noted in the introduction, for the last two decades there has been active research on 
 finding a way to build MNPDVR basis set on curved manifolds. Recently Littlejohn {\it et al} \cite{Littlejohn}; 
 Dawes and Carrington \cite{Dawes}; Yu \cite{Yu}; and Degani and Tennor \cite{Tennor} reported some promising 
 results. Dawes and Carrington's work is based on an approximate simultaneous diagonalization of the multidimensional 
 co-ordinate operators described in $\mathbbm{M}_{c}$. Yu presented a scheme somewhat apparently related to 
 that reported in ref \cite{Dawes}. The basis of the work presented in ref \cite{Tennor} is that introduced 
 in ref \cite{Littlejohn} by Littlejohn {\it et al}, which is the work we intend to look into (albeit briefly) here. 
 
 \section{Reproducing Kernel functions and MNPDVR basis sets}
 Littlejohn {\it et al} took a novel approach to solving the ongoing problem of building MNPDVR basis sets. 
 The authors `axiomize' the known properties of a one-dimensional DVR basis set in the hope of deducing some 
 underlying mathematical structures of the DVR concept with which one can formulate a general framework for 
 constructing higher dimensional DVR basis sets. The essence of their work is given in the following definition. 
 (Note that in the following discussion --- for the sake of clarity --- $x$ denotes $x^{(1:d)}$, unless stated otherwise.) 
 
 {\it {\bf Definition 1 (adapted from ref \cite{Littlejohn,Little1,Little2}):} 
 Let $\mathbbm{M}_{c}$ and $\mathscr{H}$ be as defined before; $\mathcal{P}$ a projection 
 operator on $\mathscr{H}$; $\mathscr{S}=\mathcal{P}\mathscr{H}$, the space upon which $\mathcal{P}$ 
 projects, with a dimension of $N$. Then $\{\Delta_{\alpha}(x)=\mathcal{P}\delta(x-x_{\alpha})\}$ 
 and $\{x_{\alpha}\}$, where $\alpha=1,..,N$, is a DVR set.}
 
 To our knowledge, apart from two special cases \cite{Little1,Little2}, the authors have not 
 managed to achieve their ultimate goal of generating a general framework upon which MNPDVR 
 basis sets can be built. Before we come to the central idea presented in this article, it is 
 perhaps useful to recall that the so-called Dirac's $\delta$ function (or more precisely tempered distribution) 
 is not an element of $\mathscr{H}$ \cite{Natingale,Yoshika,Geroch} -- it is a functional that acts on $\mathscr{H}$ and
 resides in a dual space that is defined with reference to $\mathscr{H}$. We think the beauty 
 of {\bf Def.1} becomes clear if one replaces $\mathscr{H}$ in the definition with a `Rigged Hilbert Space'. 
 With this minor amendment, it is then possible to use loosely $\Delta_{\alpha}(x)=\mathcal{P}\delta(x-x_{\alpha})$ and $\Delta_{\alpha}(x)=<x|\mathcal{P}|x_{\alpha}>$ interchangeably. In the following discussion, for both pedantry and convenience, we elect to use the former which can obviously be expressed as
 \begin{eqnarray}
 \Delta_{\alpha}(x)& =& \int_{\mathbbm{M}}\mathcal{P}(x,x^{\prime\prime})\delta(x^{\prime\prime}-x_{\alpha})dx^{\prime\prime}\\
 &=&\displaystyle\sum^{N}_{j=1}\psi_{j}(x)\psi_{j}(x_{\alpha})\nonumber
 \end{eqnarray}
 where  
 $\mathcal{P}(x,x^{\prime\prime})=\displaystyle\sum^{N}_{j=1}\psi_{j}(x)\psi_{j}(x^{\prime\prime})$ \cite{qm}  

 Because of the closure property of $\{\psi_{j}(x)\}_{j=1}^{\infty}$,\\
 $\displaystyle\sum^{\infty}_{j=1}\psi_{j}(x)\psi_{j}(x_{\alpha})=\delta(x-x_{\alpha})$. 
 In other words,
 \begin{equation}
 \displaystyle\lim_{N\to\infty}\Delta_{\alpha}(x)=\displaystyle\lim_{N\to\infty}\sum^{N}_{j=1}\psi_{j}(x)\psi_{j}(x_{\alpha})=\delta(x-x_{\alpha})
 \end{equation}
 ({\it cf}. Eqs. 16 and 17 in ref. \cite{Wu}).\\
 This means that $\displaystyle\lim_{N\to\infty}\Delta_{\alpha}(x_{\alpha})=\displaystyle\lim_{N\to\infty}\sum^{N}_{j=1}|\psi_{j}(x_{\alpha})|^{2}=\infty$   
 
 Then by deduction, it can be argued that
 \begin{equation}
 \Delta_{\alpha}(x_{\alpha})=\sum^{N}_{j=1}|\psi_{j}(x_{\alpha})|^{2}<\infty 
 \end{equation}  
 which implies that 
 $\Delta_{\alpha}(x)\in\mathscr{D}~(\subset\mathscr{H})={\it span}\{\psi_{j}(x)\}^{N}_{j=1}$ even though 
 $\displaystyle\lim_{N\to\infty}\Delta_{\alpha}(x)\notin \mathscr{H}$, Eq. 5. By definition \cite{Littlejohn,CPC} $x_{\alpha}$ is a DVR point whose corresponding DVR function is given by $\Delta_{\alpha}(x)= \displaystyle\sum^{N}_{j=1}\psi_{j}(x)\psi_{j}(x_{\alpha})$ where $R_{\alpha{\it j}}=\psi_{j}(x_{\alpha})$ ({\it cf.} Eq.3), {\it i.e.} ${\bf R}^{\dagger}{\bf R} ( = {\bf R}^{-1}{\bf R}) =\mathbbm{1}_{N}$; this means $\Delta_{\alpha}(x_{\beta})=\delta_{\alpha\beta}=\displaystyle\sum^{N}_{j=1}\psi_{j}(x_{\beta})\psi_{j}(x_{\alpha})$ when $x_{\alpha}$ and $x_{\beta}$ are DVR points. Since ${\bf R}$ is unitary, then $\mathscr{D}$ is in the span of the set $\{\Delta_{\alpha}(x)\}_{\alpha=1}^{N}$ as well. (Note that we used $R_{\alpha{\it j}}=\psi_{j}(x_{\alpha})$ instead of $R_{\alpha{\it j}}=c_{\alpha}\psi_{j}(x_{\alpha})$ as $c_{\alpha}$ can be absorbed into $\psi_{j}(x_{\alpha})$; this is also an implicit assumption made in the rest of the paper.)  
 
Now we come to the nub of this paper. Using the facts given in the preceding paragraphs,  Eq. 4 
 can obviously be rewritten as 
 \begin{eqnarray}
  \Delta_{\alpha}(x)&=&\sum^{N}_{\beta=1}({\bf R}^{-1})_{\alpha\beta}\sum^{N}_{j=1}\psi_{j}(x)\psi_{j}(x_{\beta})\\\nonumber  
  &=&\sum^{N}_{\beta=1}({\bf R}^{-1})_{\alpha\beta}\sum^{N}_{j=1}k(x,x_{\beta})\nonumber   
 \end{eqnarray} 
where
 \begin{equation}
  k(x,x_{\alpha})=\sum^{N}_{j=1}\psi_{j}(x)\psi_{j}(x_{\alpha})
 \end{equation} 
 From Eqs. 7 and 8, it is clear that $k(x,x_{\alpha})$ is symmetric and strictly positive definite \cite{Aronszajn}. 
 According to Eq. 6, $\displaystyle\sup_{x_{\alpha}\in{M}}k(x_{\alpha},x_{\alpha})<\infty$ which 
 implies $k$ is a valid representation of the inner-product in the span $\{\psi_{j}(x)\}_{j=1}^{N}$.
 In other words, $k(x,x_{\alpha})$ is a reproducing kernel (positive definite) function  \cite{Aronszajn,Moore,alis,Christina}. All these points indicate that generating a DVR function, $\Delta_{\alpha}(x)$, via $\mathcal{P}\delta(x-x_{\alpha})$ can be viewed as finding a reproducing kernel function $k(x,x_{\alpha})$ defined over $\mathbbm{M}$, such that equations 7 holds. 
 
 First a brief description of the reproducing kernel theory is given in the following paragraph in 
 the form of a definition. For more detailed discussions on the subject, the reader is referred to 
 the seminal papers of N. Aronszajn \cite{Aronszajn} and E. H. Moore \cite{Moore}. More descriptions 
 on the subject, both on its application and theory, may be found in Refs. \cite{alis,Christina,kernelqn1,
 field,Rabitz}.
  
 {\it {\bf Definition 2 (adapted from ref. \cite{cambla}):} Let $\mathbbm{X}$ be any domain and let\\ 
 $\mathit{q}$({\it s},{\it r}) ---~{\it s} and {\it r}~$\in\mathbbm{X}$ --- be symmetric and positive 
 definite function if for any $M\in\mathbbm{N}$, and any {\it r}$_{1}$, {\it r}$_{2}$,....,
 {\it r}$_{M}\in\mathbbm{X}$, the {\bf H}$_{M \times M}$ matrix is positive definite; with the 
 ({\it l,m})$^{th}$ element given by $\mathit{q}({\it s}_{l},{\it r}_{m}$). Then $\mathit{q}({\it s},{\it r})$
 is a reproducing kernel function, with which a unique RKHS $\mathcal{H}_{\mathit{h}}$ is associated, 
 and can be described as having the representation 
 \begin{equation}
  \mathit{q}({\it s},{\it r}) =\sum^{N}_{j=1}\zeta_j({\it s})\zeta_{j}({\it r})
 \end{equation}
  where $\zeta_j({.})$ is defined on $\mathbbm{X}$. A function is a reproducing kernel in the sense 
  of $<f(.),\mathit{q}(.,{\it r})> = f({\it r}), \forall f(.)\in\mathcal{H}_{\mathit{h}}$.} 
  
  The last statement means $\mathit{q}({\it s},{\it r})$ is a representer of a bounded linear functional 
  which acts on RKHS (here onwards denoted by $\mathcal{H}_{\mathit{q}}$) \cite{Aronszajn,alis}. It also 
  means that the set $\mathit{q}({\it s},{\it r}_{i}), i=1,2,...,M$ can span $\mathcal{H}_{\mathit{q}}$ \cite{Aronszajn}.
  It should also be noted that just $\mathit{q}({\it s},{\it r})$ being a positive definite function on $\mathbbm{X}$ is 
  a sufficient condition for that function to be a reproducing kernel \cite{Aronszajn,alis}, an observation the importance
  of which becomes clear later in the article. (Without lost of generality in the following discussions, 
  $M$ is set to $N$ --- and it is worth noting that $M$ can not be $>N$ \cite{zhou}. ) 
  
  As stated in the definition, a function is positive definite over the given domain if ${\bf a}^{T}{\bf H}_{N\times N}{\bf a}=\displaystyle\sum^{N}_{l,m }a_{l}a_{m}\mathit{q}(s_{l},r_{m})\geq0$, where $\forall~a_{l}$ and $a_{m}\in\mathcal{R}\backslash 0$. 
  
  For the purpose of generating MNPDVR basis sets, another equally crucial property of $\mathit{q}$ is that if (say) $\mathit{t}_{1}$, $\mathit{t}_{2}$, and so on, are reproducing kernel functions defined over a manifold, then using these kernel functions we may be able to design other reproducing 
kernels defined over the same domain \cite{Christina,cambla}. 
 
  Clearly, based on the above definition, $\mathit{k}(x,x_{\alpha})$ (see Eq. 8) is 
  a reproducing kernel function \cite{Aronszajn,mercer} whose associated $\mathcal{H}_{\mathit{h}}$ can be 
  viewed as $\mathscr{D}$ (cf. $\mathscr{S}$ noted in {\bf Def.1}) provided that $N<\infty$. 
  In the following discussion we demonstrate how a DVR function can be generated from a positive 
  definite function. 
 
  If {\bf Q} is strictly positive definite\footnote{Recall that if $\mathit{q}(x,x_{\alpha})$ is not 
  strictly positive definite, it can be turned into one, see 
  Refs.\cite{Christina,cambla}}, {\it i.e.}, $\displaystyle\sum^{N}_{l,m }a_{l}a_{m}\mathit{q}(s_{l},r_{m})>0$, 
  then ${\bf Q}^{-1}{\bf Q} = \mathbbm{1}_{N}$. In other words,  
  \begin{equation}
   ({\bf Q}^{-1}{\bf Q})_{\alpha\beta}=\delta_{\alpha\beta}=\sum_{\beta=1}^{N}({\bf Q}^{-1})_{\alpha\beta}\mathit{q}(x_{\beta},x_{\alpha}), ~s_{l}=x_{\beta};~r_{m}= x_{\alpha}
   \end{equation}
   {\it i.e.},  
   \begin{equation}
    \upsilon_{\alpha}(x)= \sum_{\beta=1}^{N}({\bf Q})^{-1}_{\alpha\beta}\mathit{q}(x_{\beta},x) 
   \end{equation}
   is a function that $\displaystyle\upsilon_{\alpha}(x_{\beta})=\delta_{\alpha\beta}$, with $\alpha$ ={\it 1,..., $N$}.
   
   Of course, as $\displaystyle\upsilon_{\alpha}(x)$ is a linear combination of a set of reproducing 
   kernels, $\displaystyle\upsilon_{\alpha}(x)$ itself is a reproducing kernel function \cite{Christina,cambla}
   whose relation to $\Delta_{\alpha}(x)$ can be readily shown as follows
  \begin{eqnarray}
  \displaystyle \upsilon_{\alpha}(x)&=& \sum_{\beta=1}^{N}({\bf Q})^{-1}_{\alpha\beta}\mathit{q}(x_{\beta},x)
   =\sum_{\beta=1}^{N}({\bf Q})^{-1}_{\alpha\beta}\sum_{j=1}^{N}\zeta_{j}(x_{\beta})\zeta_{j}(x)\\\nonumber
  &=&\sum_{j=1}^{N}\bigg[\sum_{\beta=1}^{N}({\bf Q})^{-1}_{\alpha\beta}\zeta_{j}(x_{\beta})\bigg]\zeta_{j}(x)
  =\sum_{j=1}^{N}c_{\alpha}\zeta_{j}(x_{\alpha})\zeta_{j}(x)\\
  &=&\sum_{j=1}^{N}\psi_{j}(x_{\alpha})\psi_{j}(x)\\
  &=&\mathcal{P}[\delta(x-x_{\alpha})] = \Delta_{\alpha}(x)
  \end{eqnarray}
  ; which is what we set out to show. (Note that $\psi_{j}(.)=\sqrt(c_{\alpha})\zeta_{j}(.)$. )
  
 Now we briefly analyse the implications of Eqs. 12--14. {\it The equivalence of 
 Eqs. 12--14 implies that the knowledge of a positive definite function defined on 
 the specified domain is sufficient to construct a DVR basis set. In other words, 
 it is not necessary to know that set $\{\zeta_{j}(x)\}_{j=1}^{N}$ that yields 
 $\mathit{q}(x,x_{\alpha})$. A function that represents a valid inner-product in 
 the span $\{\zeta_{j}(x)\}_{j=1}^{N}$, {\it i.e.}, the appropriate $\mathit{q}(x,x_{\alpha})$,
 is all that is required \cite{Vapnick}}.
 
 From {\bf Def.2}, Eqs. 12--14 are true for both $\mathbbm{M}$ and $\mathbbm{M}_{c}$. 
 One inference that can be made from Eqs. 12--14 is that the commonly employed DVR schemes, 
 such the {\it Sinc}-DVR of Colbert and Miller \cite{Miller} and the `conventional' DVR of 
 Light and co-workers \cite{Light} are different instances of the proposed scheme. Furthermore 
 when the domain over which $\mathit{q}(x,x_{\alpha})$ is defined is one-dimensional, the Lagrangian 
 functions of Baye and co-workers (as defined in Ref. \cite{Baye}) are similar to $\mathit{h}(x,x_{\alpha})$.
 
 The equations, Eqs. 12--14, also imply that one now has control over the choice of the DVR points 
 unlike the currently used conventional DVR techniques where the global basis functions determine where on 
 $\mathbbm{M}/\mathbbm{M}_{c}$ the DVR points appear. 
  
 {\it Possible simple reproducing kernel functions:} 
 The class of parameterisable functions \footnote{with parametric values that are different from 
 their limiting values of the parameters}, $\displaystyle\frac{a}{\pi}\displaystyle\lim_{a\to\infty}\displaystyle\frac{a\sin(x-x_{\alpha})}{(x-x_{\alpha})}$,
 $\displaystyle\frac{1}{\sqrt{\pi}}\displaystyle\lim_{\epsilon\to 0}\displaystyle\frac{1}{\sqrt{\pi}}\exp^{-(\frac{x-x_{\alpha}}{\epsilon})^{2}}$, $\displaystyle\frac{1}{\pi}\displaystyle\lim_{\epsilon\to 0}\displaystyle\frac{\epsilon}{(x-x_{\alpha})^{2}+\epsilon^{2}}$ (with $a$ and $\epsilon~(\in\mathscr{R})>0$), 
 that are widely used in physics to approximate/represent the Dirac's delta function $\delta(x^{(i)}-x^{(i)}_{\alpha})$ on 
 $\mathbbm{M/M_{c}}$ provides a good example of positive definite functions.
 In the cases of $a<\infty$ and $\epsilon>0$, these functions are valid reproducing 
 kernels. In fact, the Lorentzian $\mathit{h}(x,x_{\alpha})=\displaystyle\frac{1}{\pi}\displaystyle\lim_{\epsilon\to 0}\displaystyle\frac{\epsilon}{(x-x_{\alpha})^{2}+\epsilon^{2}}$, 
 with $\epsilon >0$, can be modified to $\mathit{q}(x,x_{\alpha})=\displaystyle\frac{1}{\gamma(x-x_{\alpha})^{2}+1}$, 
 where $\gamma(\in\mathscr{R})>0$ --- a function that is a fine positive definite function on any 
 inner-product manifold \cite{Schoenberg1,cheney} of which, of course, $\mathbbm{M}$ and $\mathbbm{M}_{c}$
 are examples. In the high dimensional case, $x$ becomes ${x}^{(i=1,....d)}$. 
  
 In the literature on DVR basis sets a 2-dimensional surface $\mathcal{S}^{2}$ of 
 a unit sphere is cited frequently as an example of a curved manifold \cite{Demonic,Corey,Hank}. 
 In the reproducing kernel function context, 
 the issue of what kind of functions defined on $\mathcal{S}^{\infty}$ are positive definite 
 was comprehensively addressed in the early 1940's by I.~J.~Schoenberg \cite{Schoenberg1}. 
 Recently it was demonstrated that \cite{Meng}, for example, 
 $\displaystyle\sum_{l=0}^{\infty}a_{l}P_{l}(\mathit{p}.\mathit{p}^{\prime})$ is strictly positive definite 
 on $\mathcal{S}^{2}$ if $a_{l}\geq 0$, $\displaystyle\sum_{l=0}^{\infty}a_{l}P_{l}(1)<\infty$ and the set 
 $\{l: a_{l}>0\}$ contains infinitely many odd and ifinitely many even integers. $\mathit{p}.\mathit{p}^{\prime} = \cos\theta\cos\theta_{\alpha}+\sin\theta\sin\theta_{\alpha}\cos(\phi-\phi_{\alpha})$;
 $P_{l}$ are Legendre polynomials; $0\leq\theta\leq 2\pi; $ $0\leq\phi\leq\pi$; 
  $\mathit{p}=(\theta,\phi)$; $\mathit{p^{\prime}}=(\theta_{\alpha},\phi_{\alpha})$.
 
 This means 
\begin{equation}
 \displaystyle\lim_{N\to \infty}\frac{1}{4\pi}\sum_{l=0}^{N}(2l+1)P_{l}(\mathit{p}.\mathit{p}^{\prime})\rightarrow\delta(p-p^{\prime}) 
\end{equation}
 is a strictly positive definite function on the 2-dimensional spherical surface if $N<\infty$, while 
 $ P_{l}(\mathit{p}.\mathit{p}^{\prime})=\frac{4\pi}{2l+1}\displaystyle\sum_{m=-l}^{l}Y^{m}_{l}(\mathit{p})(Y_{l}^{m}(\mathit{p}^{\prime}))^{*}$
 (where $Y^{m}_{l}$ are spherical harmonics \cite{Potts}) is a reproducing kernel function on the 
 surface of the unit sphere. ($\frac{3}{2} - \cos(\mathit{p}.\mathit{p}^{\prime}))^{-\frac{1}{2}}$, 
 $e^{(\mathit{p}.\mathit{p}^{\prime})}$, to name but few, are other (strictly) positive definite 
 functions on $\mathcal{S}^{\infty}$.   
 
 In passing, the practical issues such as the inversion of {\bf Q} when $\mathit{d}$(or $\bar{\mathit{d}}$) 
  is large, the calculation of the matrix elements of {\bf Q} on on multi-dimensional flat and curved manifolds,
  etc, will be discussed and elaborated upon in detail elsewhere. In the following subsection we 
  illustrate (and demonstrate) the proposed scheme via solving the eigen-problem for a bound system that is described 
  on `one-dimensional manifold'. 
   
 \subsection{One Dimensional Test Case:}    
 The one-dimensional Hamiltonian operator is in the form
 \begin{equation}
 {\bf H}(x) = -\frac{1}{2}\frac{d^2}{dx^2} + V(x)
 \end{equation}
 
 In the basis set $\{\upsilon_{\alpha}(x)\}^{N}_{\alpha=1}$, the Hamiltonian 
 and (possible) overlap matrix elements ($H_{\alpha\alpha^{'}}$ and 
 $B_{\alpha\alpha^{'}}$, respectively) can be written as
 \begin{eqnarray}
 H_{\alpha^{'}\alpha}&=& T_{\alpha^{'}\alpha} +  V_{\alpha^{'}\alpha}\\
 B_{\alpha^{'}\alpha}&=& <\upsilon_{\alpha^{'}}|\upsilon_{\alpha}>\\
 \end{eqnarray}
 with 
 \begin{eqnarray}
 <\upsilon_{\alpha^{'}}|\upsilon_{\alpha}>&=& \sum^{N}_{\beta^{'},\beta}({\bf Q}^{-1})^{T}_{\alpha^{'}\beta^{'}}{\bf Q}^{-1}_{\alpha\beta}C_{\beta^{'}\beta};,~C_{\beta^{'}\beta}=<q(x_{\beta^{'}},x)|q(x_{\beta},x)> \\
 T_{\alpha^{'}\alpha}&=& \sum^{N}_{\beta^{'},\beta}({\bf Q}^{-1})^{T}_{\alpha^{'}\beta^{'}}{\bf Q}^{-1}_{\alpha\beta}<q(x_{\beta^{'}},x)|\frac{d^2}{dx^2}|q(x_{\beta},x)> \\
 \end{eqnarray}
 Since $\upsilon_{\alpha}(x)$ are the DVR functions, which are not 
 necessarily orthonormal, then
 \begin{eqnarray}
 V_{\alpha^{'}\alpha}&=& <\upsilon_{\alpha^{'}}(x)|V(x)|\upsilon_{\alpha}(x)>\\
 &=&\sum^{N}_{\beta^{'},\beta}({\bf Q}^{-1})^{T}_{\alpha^{'}\beta^{'}}{\bf Q}^{-1}_{\alpha\beta}<q(x_{\beta^{'}},x)|V(x)|q(x_{\beta},x)>\\ 
 &\approx& V(x_{\alpha})\sum^{N}_{\beta^{'},\beta}({\bf Q}^{-1})^{T}_{\alpha^{'}\beta^{'}}{\bf Q}^{-1}_{\alpha\beta}<q(x_{\beta^{'}},x)|q(x_{\beta},x)>\\
 &=&V(x_{\alpha})\sum^{N}_{\beta^{'},\beta}({\bf Q}^{-1})^{T}_{\alpha^{'}\beta^{'}}{\bf Q}^{-1}_{\alpha\beta}C_{\beta^{'}\beta} = V(x_{\alpha})<\upsilon_{\alpha^{'}}|\upsilon_{\alpha}>\\\nonumber  
\end{eqnarray}
 
 {\it i.e.}, in matrix form the Hamiltonian operator is given as 
 \begin{equation}
 {\bf H} = ({\bf Q}^{-1})^{T}{\bf T}{\bf Q}^{-1} + {\bf V}{\bf B} 
 \end{equation}

 Thus yielding the solution of the eigen-problem amounts to determing 
 the eigenpairs ($E_{n},{\bf u}_{n})$ of a general eigenproblem
 \begin{equation}
 \bigg[({\bf Q}^{-1})^{T} {\bf T}{\bf Q}^{-1} + {\bf V}{\bf B}\bigg]{\bf u}_{n}= E_{n}{\bf B}{\bf u}_{n}
 \end{equation}

 As a concrete example we chose a Gaussian function $\mathit{q}(x,x_{\alpha})=\displaystyle\bigg(\frac{2A}{\pi}\bigg)^{1/4}\exp\bigg[-A(x -x_{\alpha})^{2}\bigg]$ as 
 \cite{Hamil} a reproducing kernel, which is defined over $\mathscr{R}$. $A$ is a parameter that is tuned to render the kernel 
 strictly positive definite in the interval [$x_{min}$,$x_{max}$] on which the potential function is defined. Recall that 
 when $\mathit{q}(x,x_{\alpha})$ is strictly positive definite {\bf Q} is invertible, which in turn implies the functions 
 $\{q(x_{\alpha},x)\}$ are linearly independent over the chosen coordinate range. In other words, the small eigenvalues of 
 {\bf B} are not too small \cite{Hamil} to create numerical problems (and references therein).  
 
 When $\mathit{q}(x,x_{\alpha})$ is Gaussian, the evaluation of $<q(x_{\beta^{'}},x)|\frac{d^2}{dx^2}|q(x_{\beta},x)>$ 
 and $C_{\beta^{'}\beta}=<q(x_{\beta^{'}},x)|q(x_{\beta},x)>$ is straightforward, see Eqs. 9--11 of Ref. \cite{Hamil}.

 In this test case, the potential function in the Hamiltonian operator was the Morse 
 potential, $v(x) = D(e^{-2\xi x } - 2e^{-\xi x} )$, that was considered in Refs. \cite{Hamil, Heller}. 
 With $D=12$ and $\xi=0.2041241$, this potential function supports 24 bound eigenstates.  
 
 Finally the range $[x_{max}, x_{min}]$ was set to [--4.,~45.]. In the example, $x_{\alpha} = 
 x_{min} + \frac{x_{max} - x_{min}}{L}*(\alpha-1)$, {\it i=1,....L}. The $x_{\alpha}$ points ($L$ of them) 
 were used to generate a positive definite {\bf Q} matrix of $L\times\L$. {\bf Q} was then inverted 
 employing a LU decomposition algorithm. The chosen points became the DVR points while their 
 corresponding functions were obtained using Eq. 12. 
 
 With $L$ and $A$ set to 90 and 0.9, respectively,  the matrix elements $V_{\alpha^{'}\alpha}$ were 
 calculated using Eq. 27.  Then Eq. 29 was solved by employing {\it DSYGV}, a {\it LAPACK} 
 routine \cite{lapack}. The yielded 24 bound eigenvalues are as shown in Column 3 of the table. 
 Since strictly speaking the potential operator is not diagonal when expressed in DVR basis 
 functions \cite{Light,Harris,szlay}, the above calculation was repeated. In this time round, 
 the potential matrix elements were computed exactly using {\it Mathematica} \cite{mathematics}. 
 Self-evidently, the two results are in excellent agreement,Column 4. seemingly this confirms the 
 validity of the proposed scheme, {\it albeit} in one-dimensional problems. Note that the above 
 discussion is valid whether the coordinate domain is flat or curved. In the presented results, 
 the coordinate domain was considered flat. Thus, $(x -x_{\alpha})^{2}$ in $\mathit{q}(x,x_{\alpha})=\displaystyle\bigg(\frac{2A}{\pi}\bigg)^{1/4}\exp\bigg[-A(x -x_{\alpha})^{2}\bigg]$
 was a simple Euclidean distance. 
 
 The analytic solutions of the Hamiltonian problem are as given in Column 2. 
 As Column 5 illustrates, the accuracy of the computed lower eigenvalues is 
 good compared to that of the computed higher eigenvalues. As explained by 
 Hamilton and Light \cite{Hamil}, this can be addressed by employing unequally 
 distributed Gaussian functions -- instead of the equally distributed Gaussian 
 functions that we chose for clarity, to demonstrate the proposed algorithm. 
 \begin{table}[!h]
 \caption{Comparison of errors of the calculated eigenvalues for 
 Morse Potential. Exact: Analytic eigenvalues; RKSDVR= eigenvalues 
 computed with $V_{\alpha^{'}\alpha}$ computed via Eq. 26; RKS eigenvalues 
 computed with $V_{\alpha^{'}\alpha}$ computed exactly, Eq. 24.}
 \begin{tabular}{|c|c|c|c|c|}
 \hline
  State Number& Exact &RKSDVR & RKSDVR-RKS& Exact-RKSDVR \\\hline
     0&     -11.5052&     -11.5052&         0.000000&           3.5150666E-07\\
     1&     -10.5469&     -10.5469&         0.000000&           1.0008970E-06\\
     2&      -9.6302&      -9.6302&         0.000000&           1.6017819E-06\\
     3&      -8.7552&      -8.7552&         0.000000&           2.1581170E-06\\
     4&      -7.9219&      -7.9219&         0.000000&           2.7113422E-06\\
     5&      -7.1302&      -7.1302&         0.000000&           3.3694112E-06\\
     6&      -6.3802&      -6.3802&         0.000000&           4.3132356E-06\\
     7&      -5.6719&      -5.6719&         0.000000&           5.8555934E-06\\
     8&      -5.0052&      -5.0052&         0.000000&           8.4396995E-06\\
     9&      -4.3802&      -4.3802&         0.000000&           1.2556453E-05\\
    10&      -3.7969&      -3.7969&         0.000000&           1.8664048E-05\\
    11&      -3.2552&      -3.2552&         0.000000&           2.7037027E-05\\
    12&      -2.7552&      -2.7552&         0.000000&           3.7583833E-05\\
    13&      -2.2969&      -2.2969&         0.000000&           4.9771613E-05\\
    14&      -1.8802&      -1.8803&         0.000000&           6.2608957E-05\\
    15&      -1.5052&      -1.5053&         0.000000&           7.4703949E-05\\
    16&      -1.1719&      -1.1720&         0.000000&           8.4475342E-05\\
    17&      -0.8802&      -0.8803&         0.000000&           9.0389728E-05\\
    18&      -0.6302&      -0.6303&         0.000000&           9.1175140E-05\\
    19&      -0.4219&      -0.4220&         0.000000&           8.6053705E-05\\
    20&      -0.2552&      -0.2553&         0.000000&           7.4869651E-05\\
    21&      -0.1302&      -0.1303&         0.000000&           5.8097745E-05\\
    22&      -0.0469&      -0.0469&         0.000000&           3.6435728E-05\\
    23&      -0.0052&      -0.0043&         0.000000&          -9.3466361E-04\\\hline
  \end{tabular}
 \end{table} 
 \section{Conclusion}
  In this article it was demonstrated that, starting off with the axiomatic definition of DVR provided 
  in ref \cite{Littlejohn}, it is possible to show the space upon which the projection operator (noted in 
  that reference) projects is a Reproducing Kernel Hilbert Space whose associated reproducing kernel 
  function can be used to generate DVR points and their corresponding DVR functions on any manifold 
  (curved or not).

  The results of the simple example that was presented as a preliminary test of 
  the new scheme seemingly confirms our mathematical prediction. Nonetheless, a more realistic 
  test of the new method will be given in subsequent publications.

{\bf Acknowledgements}

  It is a great pleasure to acknowledge Dr. J. A. Townsend for reading the manuscript. 

\end{document}